\documentclass[aps, preprint]{revtex4-1}
\usepackage{amssymb}
\usepackage{latexsym}
\usepackage{amsfonts}
\usepackage{graphics}
\usepackage{graphicx}
\usepackage{epsfig}
\usepackage{amsmath}
\usepackage{float}
\usepackage{verbatim}
\usepackage[subfigure]{graphfig}

\begin{document}

\title{Robustness of partially interdependent network formed of clustered networks}

\author{Shuai Shao,$^1$ Xuqing Huang,$^1$ H. Eugene Stanley,$^1$ and Shlomo Havlin$^{1,2}$}

\affiliation{$^1$Center for Polymer Studies and Department of Physics,
Boston University, Boston, MA 02215 USA\\
$^2$Minerva Center and Department of Physics,
Bar-Ilan University, Ramat-Gan 52900, Israel
}
\date{July 30, 2013}
\begin{abstract}
Clustering, or transitivity has been observed in real networks and its effects on their structure and function has been discussed extensively. The focus of these studies has been on clustering of single networks while the effect of clustering on the robustness of coupled networks received very little attention. Only the case of a pair of fully coupled networks with clustering has been studied recently. Here we generalize the study of clustering of a fully coupled pair of networks to the study of partially interdependent network of networks with clustering within the network components. We show both analytically and numerically, how clustering within the networks, affects the percolation properties of interdependent networks, including percolation threshold, size of giant component and critical coupling point where first order phase transition changes to second order phase transition as the coupling between the networks reduces. We study two types of clustering: one type proposed by Newman~\cite{Newman2009} where the average degree is kept constant while changing the clustering and the other proposed by Hackett $et$ $al.$~\cite{Gleeson1} where the degree distribution is kept constant. The first type of clustering is treated both analytically and numerically while the second one is treated only numerically.
\end{abstract}


\maketitle

\section{Introduction}
Complex networks is a useful approach to study structure, stability and function of complex systems~\cite{Watts1998,Albert2000,Rosato2008,Albert2002,Cohen20002001,Callaway2000,NewmanSIAM,Dorogovtsev2003,Song2005,Caldarelli2007,Cohen2010,NewmanBook,Bashan2012,Li2010,Schneider2011}. Clustering, the propensity of two neighbors of the same node  to be also neighbors of each other, has been observed in many real-world networks~\cite{Watts1998,Wasserman1994,Ravasz2003,Serrano2006}. For example, in a social network, if B and C are friends of A, they also have a high probability to be friends of each other. The average of this probability over the whole network is called clustering coefficient. Empirical studies show that in many real-world networks, e.g., the Internet, scientific collaboration networks, metabolic and protein networks and movie actors networks,  the measured clustering coefficient is of the order of $10\%$, significantly higher than that of random networks~\cite{Newman2003, Albert2002}. \\
Many computational models have been proposed to generate clustering coefficient in networks, but all limited to numerical analysis~\cite{Jin2001,Holme2002,Klemm2002,Serrano2005,Bansal2009}. Recently, Newman showed how to incorporate clustering into random graphs by extending the generating function method, a widely used analytical tool in network research~\cite{Newman2009}. He considered two properties for each node-- the single links and the triangles, and constructed a joint distribution for both.  Clustering coefficient can be tuned by changing the ratio between the average number of single links and triangles. This approach enables to evaluate analytically many properties of the resulting networks, such as component sizes, emergence and size of a giant component and other percolation properties. \\
Earlier studies on clustering have been focused on single network analysis, while most of the real-world networks interact with and depend on other networks. In 2010, Buldyrev $et$ $al.$~\cite{Buldyrev2010} developed a theoretical framework for studying percolation of two fully interdependent networks and an unusual first order percolation transition (abrupt) has been observed, which is different from the known second order phase transition (continous) in a single network. Parshani $et$ $al.$~\cite{Parshani2010} generalized the framework to partially interdependent networks and found a change from first order phase transition to second order phase transition when reducing the coupling strength below a critical value.  Recently, Huang $et$ $al.$~\cite{Xuqing2012} developed an approach for site percolation on clustered networks and studied the robustness of a pair of fully interdependent networks with clustering within each network.\\
Here, we generalize the framework of Huang $et$ $al.$~\cite{Xuqing2012} in two directions: (i) to the study of percolation of two partially interdependent networks with clustering within each network and (ii) to the study of network of networks (NON), i.e., network formed of more than two interdependent networks. We study the influence of clustering within the networks on percolation properties, such as the critical shreshold $p_c$ where the giant component collapses, sizes of the giant components $\psi_{\infty}$ and $\phi_{\infty}$ in the two networks, the critical coupling $q_c$ at which the first order phase transition changes to a second order phase transition and the dynamics of cascading failure between two clustered networks. Furthermore, the percolation of network of clustered networks is also investigated. Simulation results show very good agreement with theoretical results in all cases.\\
We also discuss in Sect. V, two joint distribution models for incorporating clustering into random graphs. (i) The model proposed by Newman~\cite{Newman2009}, where a doubly poisson distribution (see Sect. III) is assumed for the joint degree distribution and the average degree is kept constant while changing clustering. (ii) The clustering model developed by Hackett $et$ $al.$~\cite{Gleeson1}, where a different joint distribution was proposed to keep not only the average degree, but also the degree distribution constant while changing clustering. We discuss the similarity and difference of the percolation properties of networks between these two distribution models. Newman's model is studied both analytically and via simulations (Sects. III and IV) while Hackett $et$ $al.$'s model is studied only via simulations (Sect. V).
\section {The Model}
In our model, we consider two networks A and B of the same number of nodes N. Within each network, the nodes are connected with joint degree distribution $P_A(s,t)$ and $P_B(s,t)$, specifying the fraction of nodes connected to $s$ single links and $t$ triangles in network A and B, respectively~\cite{Newman2009}. The generating functions~\cite{Newman2001,Newman2002} of the joint degree distributions are 
\begin{equation}
\begin{split}
G_{A0}(x,y)&=\sum_{s,t=0}^{\infty}P_A(s,t)x^sy^t, \label{Generating}\\
G_{B0}(x,y)&=\sum_{s,t=0}^{\infty}P_B(s,t)x^sy^t.
\end{split}
\end{equation}
The conventional degree of a node is $k=s+2t$ and the conventional degree distributions of the networks are
\begin{equation}
\begin{split}
P_A(k)&=\sum_{s,t=0}^{\infty}P_A(s,t)\delta_{k,s+2t},\label{Degree}\\
P_B(k)&=\sum_{s,t=0}^{\infty}P_B(s,t)\delta_{k,s+2t}.
\end{split}
\end{equation}
The clustering coefficient is defined in~\cite{Newman2001} as
\begin{eqnarray}
c=\frac{3\times \mbox{(number of triangles in network)}}{\mbox{number of connected triples}}=\frac{3N_\Delta}{N_3},\label{Clustering}
\end{eqnarray}
where $3N_\Delta \equiv N\sum_{st}tP(s,t)$ and $N_3=N\sum_k\binom{k}{2}P(k)$. \\
We begin with an initial attack on network A by randomly removing a $(1-p)$ fraction of nodes in network A. The generating function of the resulting network is~\cite{Xuqing2012}
\begin{eqnarray}
\begin{split}
&G^{'}_{A0}(x,y)\equiv G_{A0}(x,y,p)\\
&=G_{A0}(xp+1-p,p^2y+2xp(1-p)+(1-p)^2),\label{Xuqing}
\end{split}
\end{eqnarray}
and the fraction of nodes belonging to the giant component in the remaining network is
\begin{eqnarray}
g_A(p)&=&1-G_{A0}(u,v^2,p),\label{Xuqing2}
\end{eqnarray}
where $u$,$v$ satisfy 
\begin{equation}
u=G_{Aw}(u,v^2,p),\qquad v=G_{Ar}(u,v^2,p).\label{Xuqing3}
\end{equation}
The functions $G_{Aw}(x,y,p)$ and $G_{Ar}(x,y,p)$ are defined as  
\begin{equation}
\begin{split}
G_{Aw}(x,y,p)&=\frac{1}{  \langle s^{'}\rangle}\frac{\partial G_{A0}(x,y,p)}{\partial x},\\
G_{Ar}(x,y,p)&=\frac{1}{\langle t^{'}\rangle}\frac{\partial G_{A0}(x,y,p)}{\partial y}, \label{Xuqing4}
\end{split}
\end{equation}
where $\langle s^{'}\rangle=\frac{\partial G_{A0}(x,y,p)}{\partial x}\big|_{x=1,y=1}$ and $\langle t^{'}\rangle=\frac{\partial G_{A0}(x,y,p)}{\partial y}\big|_{x=1,y=1}$. Similar equations hold for network B.\\
Next, we consider the interaction between the two clustered networks A and B~\cite{Parshani2010}. Assume a $q_A$ fraction of nodes in network A depend on nodes in network B and a $q_B$ fraction of nodes in network B depend on nodes in network A. This means that if a node in network B upon which a node in network A depends fails, the corresponding node in network A will also fail, and vice versa. Besides, we assume here that a node from one network may depend on no more than one node from the other network and if a node $i$ in network  A depends on a node $j$ in network B and $j$ depends on a node $l$ in network A, then $l=i$ (no-feedback condition~\cite{Gao2011,Gao2012}). After n steps of cascading failures, $\psi_n$ and $\phi_n$ are the fractions of nodes in the giant components of network A and network B, respectively. After the system of the two networks reaches staionarity, the sizes of giant components of the two networks can be found to be~\cite{Parshani2010}
\begin{eqnarray}
\psi_{\infty}=xg_A(x),\qquad \phi_{\infty}=yg_B(y),\label{Parshani1}
\end{eqnarray}
where the two variables $x$ and $y$ satisfy 
\begin{equation}
\begin{split}
x&=p\{1-q_A[1-g_B(y)]\},\\
y&=1-q_B[1-pg_A(x)]. \label{Parshani2}
\end{split}
\end{equation}

\section{Doubly Poisson Distribution}
As an example, consider two Erd\H{o}s-R\'{e}nyi(ER) networks~\cite{ER1,ER2,ER3} with clustering, in which the number of single links $s$ and triangles $t$ of a node obey a doubly Poisson distribution $P_{st}$=$e^{-\langle s \rangle}\frac{\langle s \rangle^s}{s!}e^{-\langle t\rangle}\frac{\langle t\rangle^t}{t!}$. Here $\langle s\rangle$ and $\langle t\rangle$ are the average numbers of single links and triangles per node, respectively~\cite{Newman2009}. Assuming for network A, $\langle s\rangle=\langle s \rangle_A$ and $\langle t\rangle=\langle t \rangle_A$, then the generating functions in Eq.~(\ref{Xuqing}) and Eq.~(\ref{Xuqing4}) become
\begin{equation}
\begin{split}
&G_{A0}(x,y,p)=G_{Aw}(x,y,p)=G_{Ar}(x,y,p)\\
&=e^{[\langle s \rangle_Ap+2p(1-p)\langle t \rangle_A](x-1)+\langle t \rangle_Ap^2(y-1)}, \label{Gawr}
\end{split}
\end{equation}
and the same holds for network B. Denote $f_A(x)=1-g_A(x)$ and $f_B(y)=1-g_B(y)$, we now have 
\begin{equation}
\begin{split}
f_A(x)&=exp\{\langle t \rangle_Ax^2(1-f_A(x))^2-\langle k \rangle_Ax(1-f_A(x))\},\\
f_B(y)&=exp\{\langle t \rangle_By^2(1-f_B(y))^2-\langle k \rangle_By(1-f_B(y))\}, \label{fab}
\end{split}
\end{equation}
where $\langle k\rangle _A$ and $\langle k \rangle_B$ are the average degrees for network A and network B, respectively ($\langle k \rangle_A=\langle s \rangle_A+2\langle t \rangle_A,$ and $\langle k \rangle_B=\langle s \rangle_B+2\langle t \rangle_B$). By combining Eqs. ~(\ref{Parshani2}) and ~(\ref{fab}) and eliminating $x$ and $y$, we obtain two transcendental equations for $f_A$ and $f_B$:
\begin{equation}
\begin{split}
f_A=&e^{\langle t \rangle_Ap^2(1-f_A)^2(1-q_Af_B)^2-\langle k \rangle_Ap(1-f_A)(1-q_Af_B)},\\
f_B=&e^{\langle t \rangle_B(1-f_B)^2\{1-q_B[1-p(1-f_A)]\}^2-\langle k \rangle_B(1-f_B)\{1-q_B[1-p(1-f_A)]\}}. \label{fafb}
\end{split}
\end{equation}
By substituting the parameter vector $(\langle k \rangle_A,\langle t \rangle_A,\langle k \rangle_B,\langle t \rangle_B,q_A,q_B,p)$, we can solve for $f_A$ and $f_B$, and thus find the size of giant components in network A, $\psi_{\infty}$, and network B, $\phi_{\infty}$. From Eqs.~(\ref{Clustering}) and~(\ref{Gawr}), the clustering coefficients in the two networks become
\begin{equation}
\begin{split}
c_A&=\frac{2\langle t \rangle_A}{\langle k \rangle_A^2+2\langle t \rangle_A},\\
c_B&=\frac{2\langle t \rangle_B}{\langle k \rangle_B^2+2\langle t \rangle_B}.
\end{split}
\end{equation}

As we fix the other parameters and increase $p$, the fraction of non-removed nodes in the initial attack, a phase transition will occur at some $p_c$ from no giant component to the existence of a giant component. As we decrease the coupling strength $q_A$ and $q_B$, the behavior of this phase transition will change from first order to second order. By adding the condition for first order phase transition~\cite{Parshani2010}, $\frac{df_B(f_A)}{df_A}\frac{df_A(f_B)}{df_B}=1$, into Eqs.~(\ref{fafb}), we can solve for $f_A=f_{A_{\uppercase\expandafter{\romannumeral1}}}$, $f_B=f_{B_{\uppercase\expandafter{\romannumeral1}}}$ and $p=p_{\uppercase\expandafter{\romannumeral1}}$, which corresponds to the scenario that the size of one or both giant components in the two networks change discontinuously from finite value to zero (first order transition). While second order phase transition (denoted by II), corresponding to the scenario that the size of one or both giant components decreases continuously to zero, is obtained by substituting $f_A\rightarrow 1$ or $f_B\rightarrow 1$ into Eqs.~(\ref{fafb}), which allows us to find $f_{A_{\uppercase\expandafter{\romannumeral2}}},f_{B_{\uppercase\expandafter{\romannumeral2}}}$ and $p_{\uppercase\expandafter{\romannumeral2}}$. The critical coupling strength $q_c$ is solved by equating the conditions for both first order phase transition and second order phase transtion.

\begin{Figure}{\label{fig1234} Size of giant components as a function of $p$ for $\langle k \rangle=\langle k \rangle _A=\langle k \rangle_B=4$, where solid lines are from theoretical predictions and symbols are from simulations with network size $N=10^5$. (a) and (b) For strong coupling ($q=0.8$), the sizes of giant components in (a) network A and (b) network B change abruptly at some critical threshold $p_c$, showing a first order phase transition behavior. (c) and (d) For weak coupling ($q=0.6$), on the contrary, the behavior is continuous, i.e., second order. Note that while (c) network A collapses (d) network B does not collapse, since the initial failures are in A and $q$ is relatively small to cause collapse of network B. Thus, a giant component of B is finite for all $p$ values.}[lables]
\graphfile*[38]{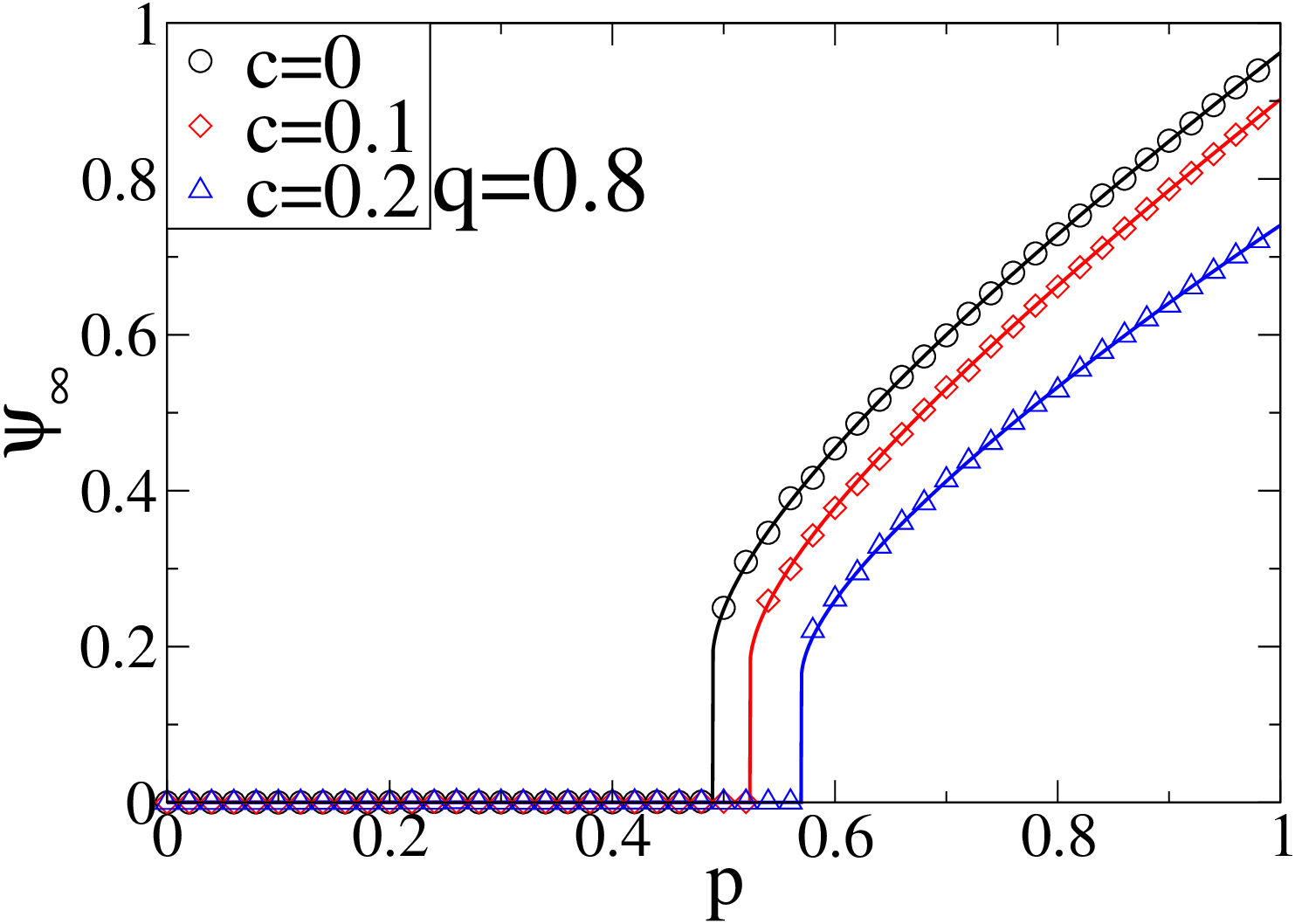}
\graphfile*[38]{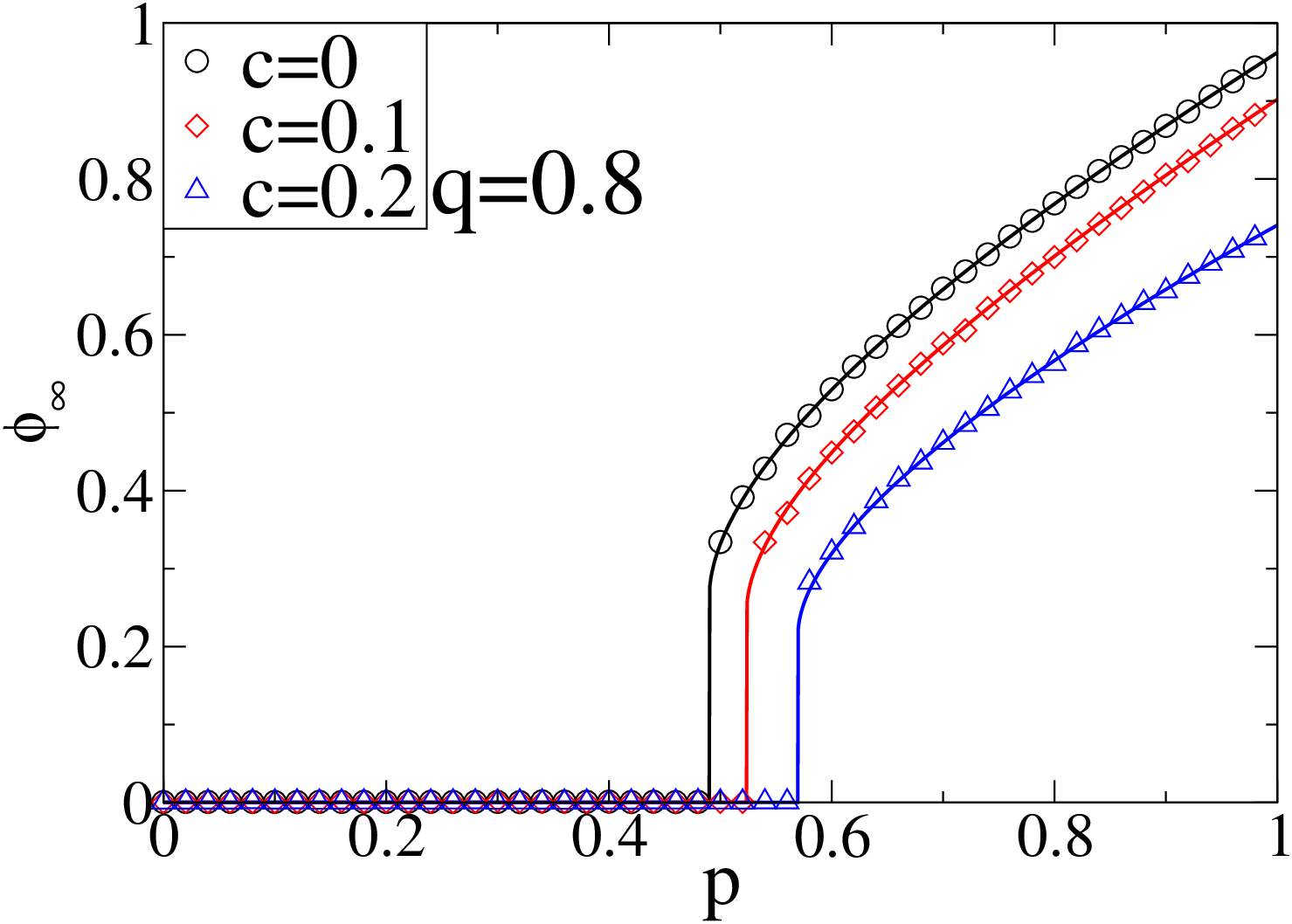}\\
\graphfile*[38]{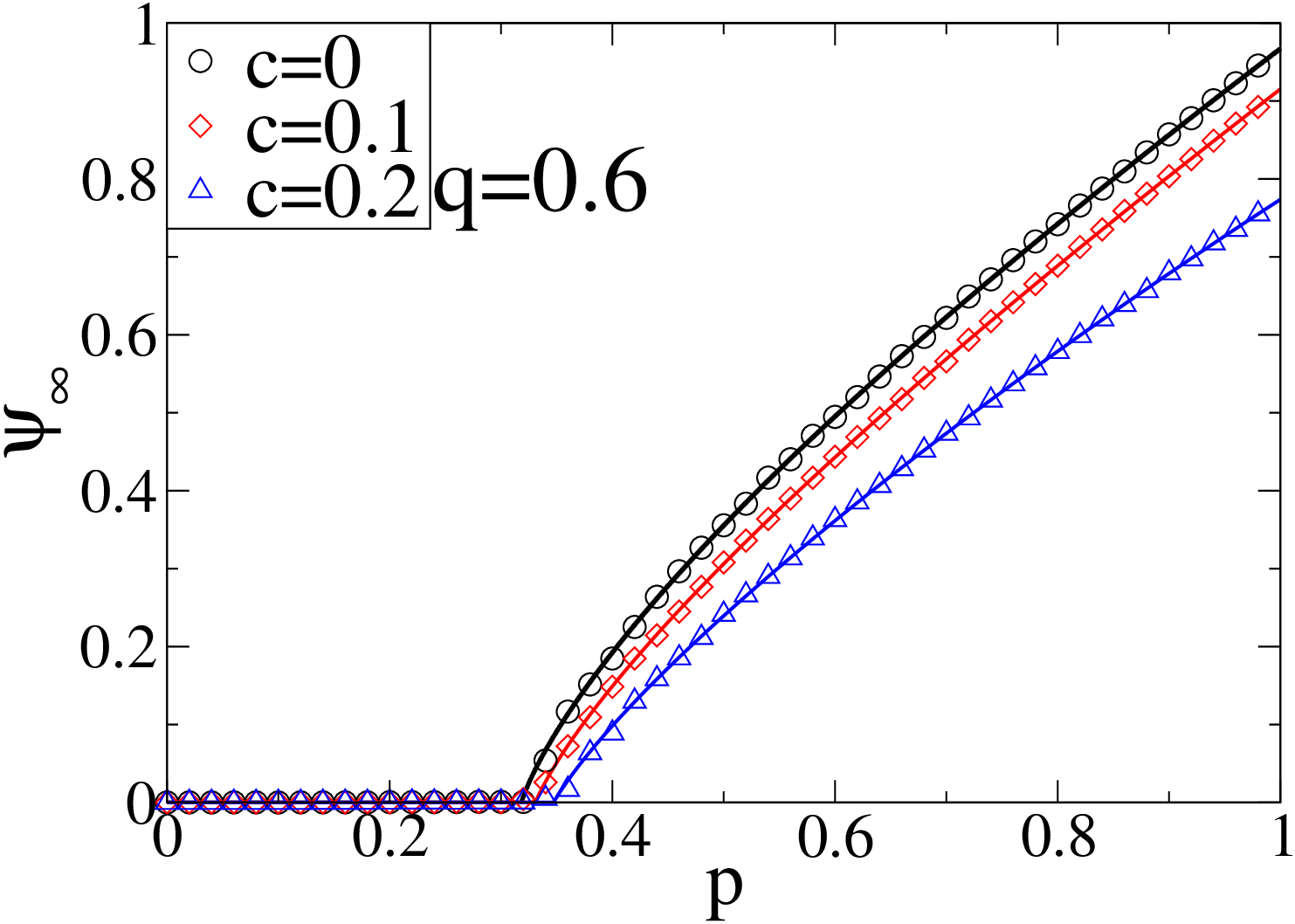}
\graphfile*[38]{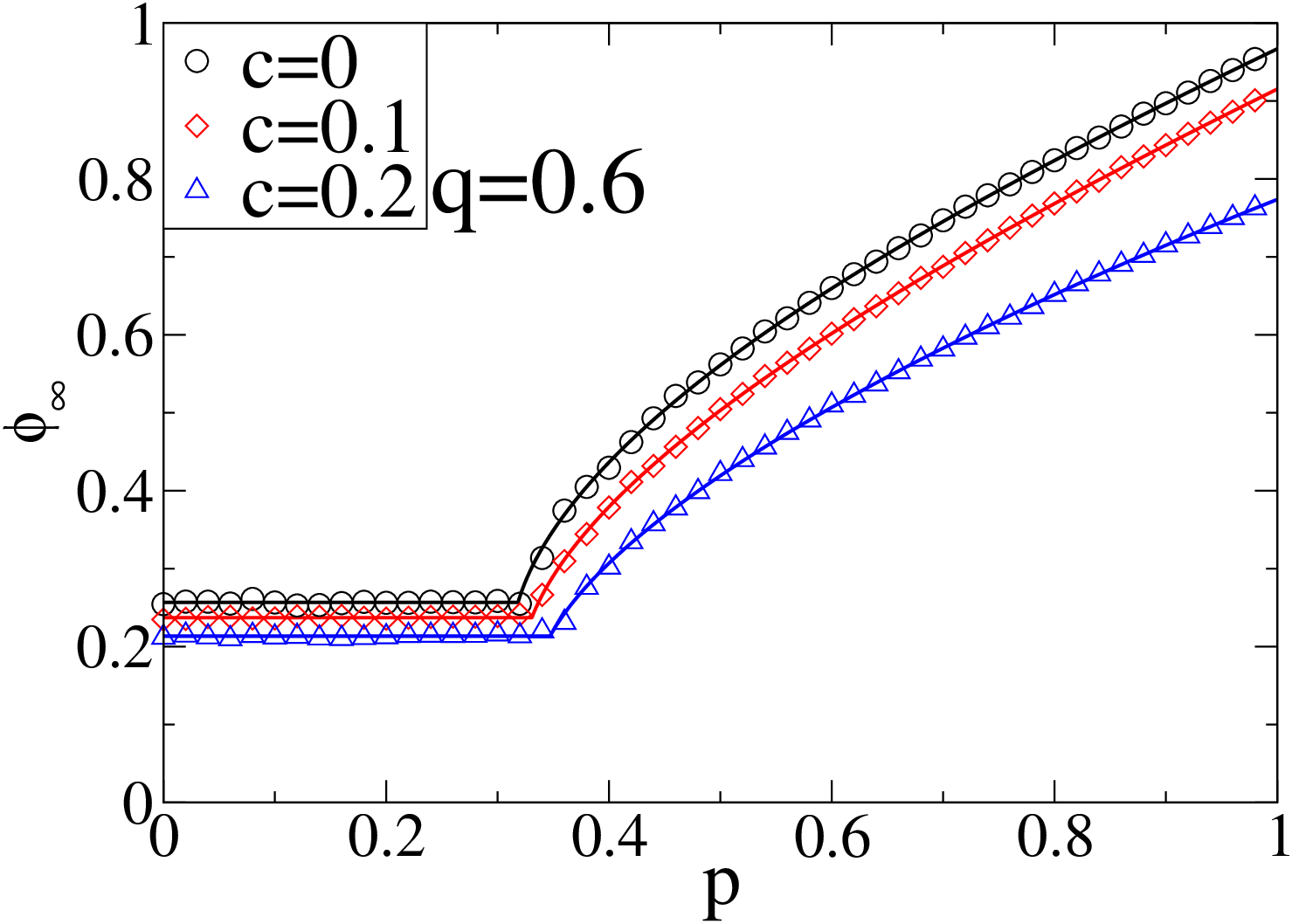}
\end{Figure}

\begin{figure}
   \includegraphics[width=0.48\textwidth, angle = 0]
   {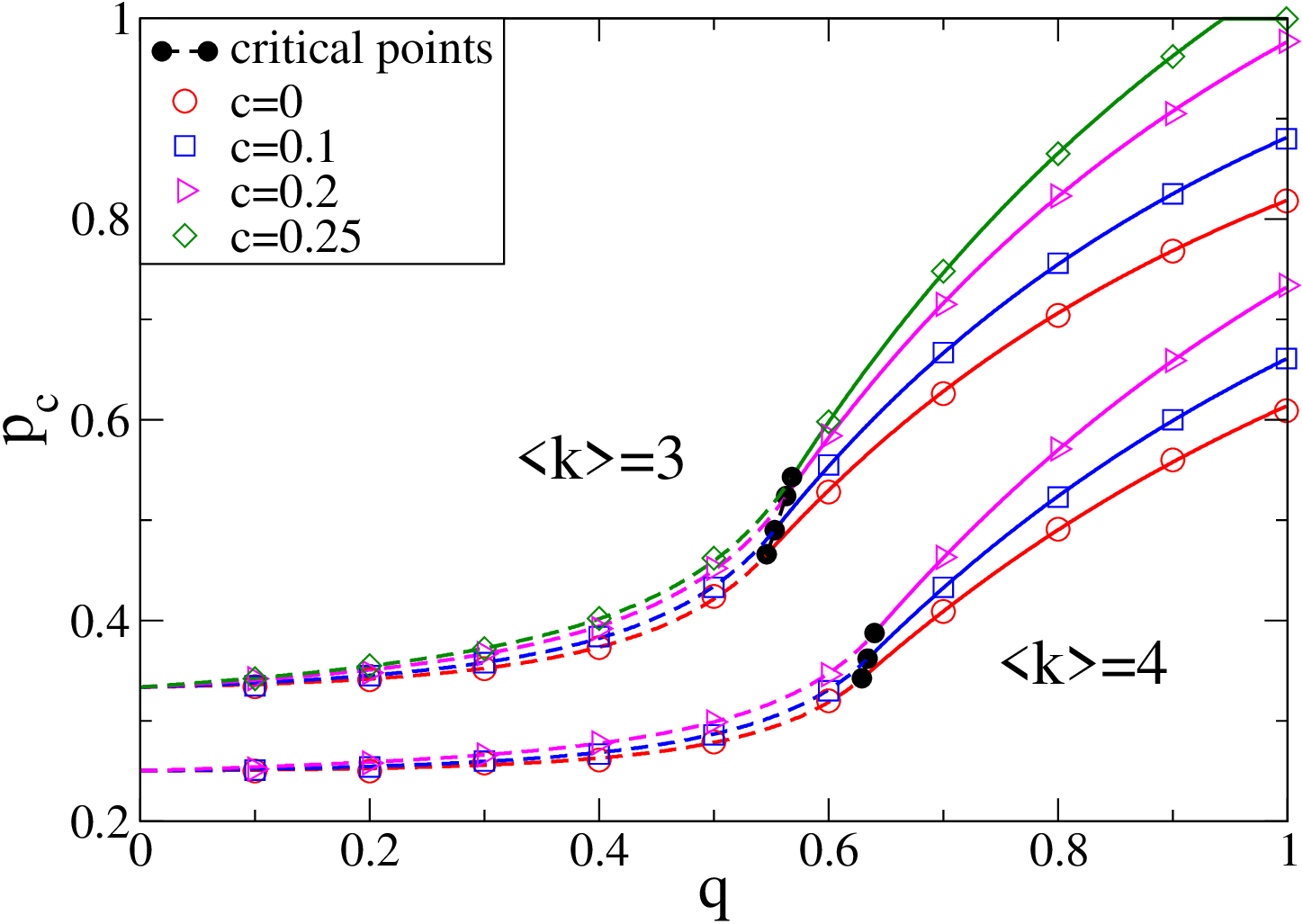}
\caption{\label{Fig2} Percolation shreshold $p_c$ as a function of interdependency strength $q$ ($q=q_A=q_B$) for $\langle k \rangle=\langle k \rangle _A=\langle k \rangle_B=3$ and 4. Clustering coefficient $c$ ($c$=$c_A$=$c_B$) ranges from 0 to 0.2 for $\langle k \rangle$=4 and from 0 to 0.25 for $\langle k \rangle$=3. For each $\langle k \rangle$ and $c$, there exists a critical point $q_c$ (full circles). Above the critical point $q_c$, the system undergoes a first order phase transition (solid lines). Below $q_c$, the system undergoes a second order transtion (dashed lines). Symbols represent simulation results and are in good agreement with theoretical predictions (solid and dashed lines). Note that for the same average degree $\langle k \rangle$, increasing clustering coefficient $c$ increases $p_c$ and yields a larger critical coupling, $q_c$.}
\end{figure}

For simplicity, we now consider the symmetrical case, $\langle k \rangle=\langle k \rangle_A=\langle k \rangle_B$ and $c=c_A=c_B$. Fig.~\ref{fig1234} shows the sizes of giant components in network A and network B for several clustering coefficients. In each graph, simulation results show excellent agreement with the theoretical results obtained from Eqs.~(\ref{fafb}). One can see that for strong coupling, as we increase the clustering coefficient, the two interdependent networks become less robust. While for weak coupling, the effect of clustering coefficient on robustness is smaller. As $q=q_A=q_B$ goes to zero, which means there is no coupling between the two networks, clustering coefficient will not affect the critical threshold for site percolation on an isolated clustered network with doubly Poisson distribution~\cite{Xuqing2012}. This can be seen more clearly from Fig.~\ref{Fig2}, which shows $p_c$ versus $q=q_A=q_B$ for different clustering coefficients for both $\langle k \rangle=3$ and $\langle k \rangle=4$. One can see that for the same coupling strength $q$,  larger clustering coefficient yields larger $p_c$, making the networks less robust. Also, the critical coupling strength $q_c$ below which the first order phase transition changes to a second order increases slightly as we increase clustering coefficient. \\

\begin{Figure}{\label{Fig3} Size of the giant component in network A ($\psi_n$) as a function of cascading failure steps $n$ for $\langle k \rangle$=4, $c$=0.2 for (a) $q=0.8$ (first order transition) and (b) $q=0.6$ (second order transition). The symbols (circles) and their connecting line are from the theoretical prediction. The other lines are several random realizations from simulations ($N=10^6$). The value of $p=0.569$ for (a) the first order phase transtion and $p=0.347$ for (b) the second order phase transition are both chosen to be just below critical thresholds obtained from theoretical predictions ($p_c=0.57$ for the first order and $p_c=0.3475$ for the second order). One can see that in both cases the agreement is perfect. However, for first order phase transition, after the plateau different realizations fluctuate.}
\graphfile*[38]{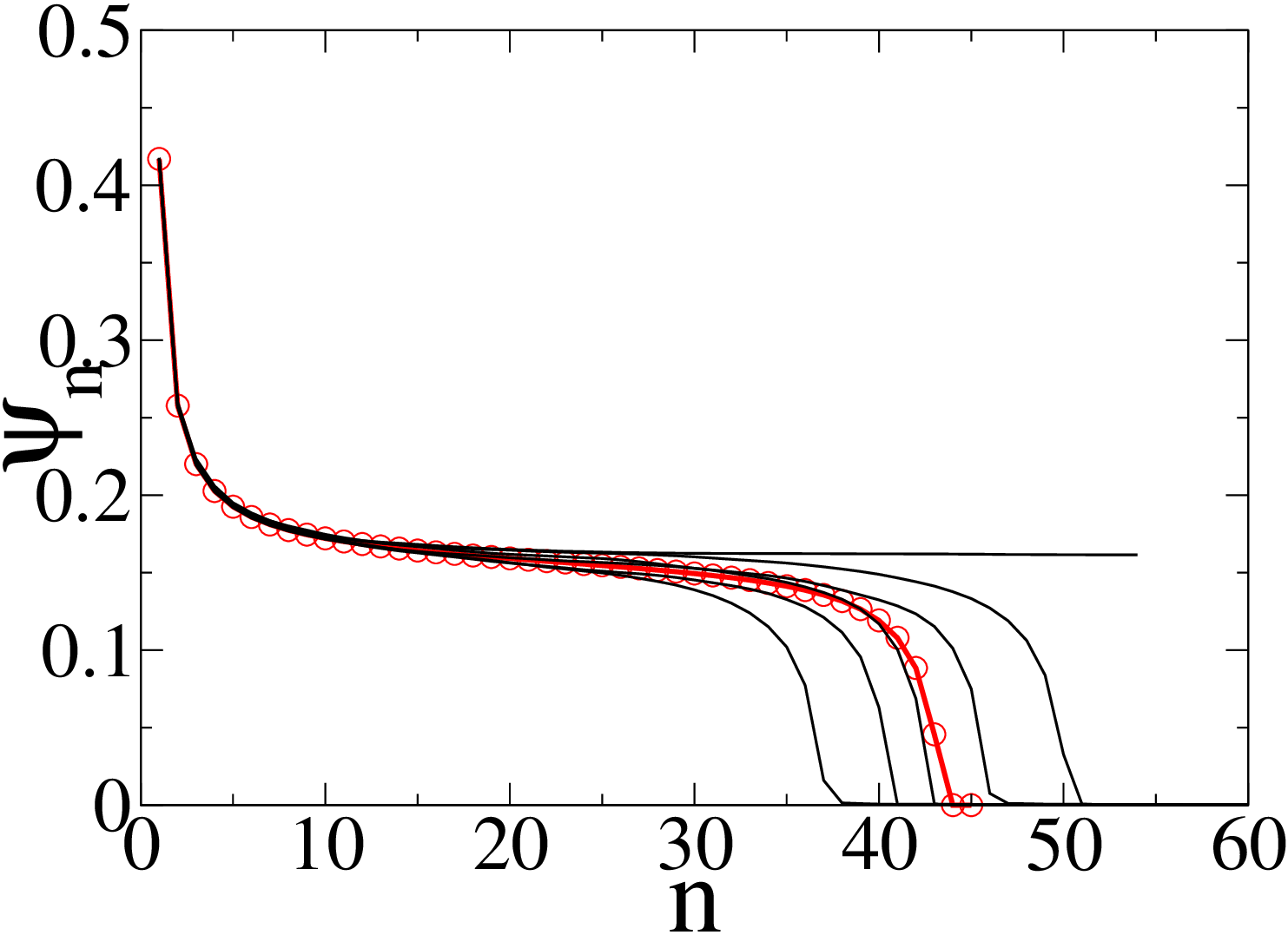}[\label{Fig333aaa}]
\graphfile*[38]{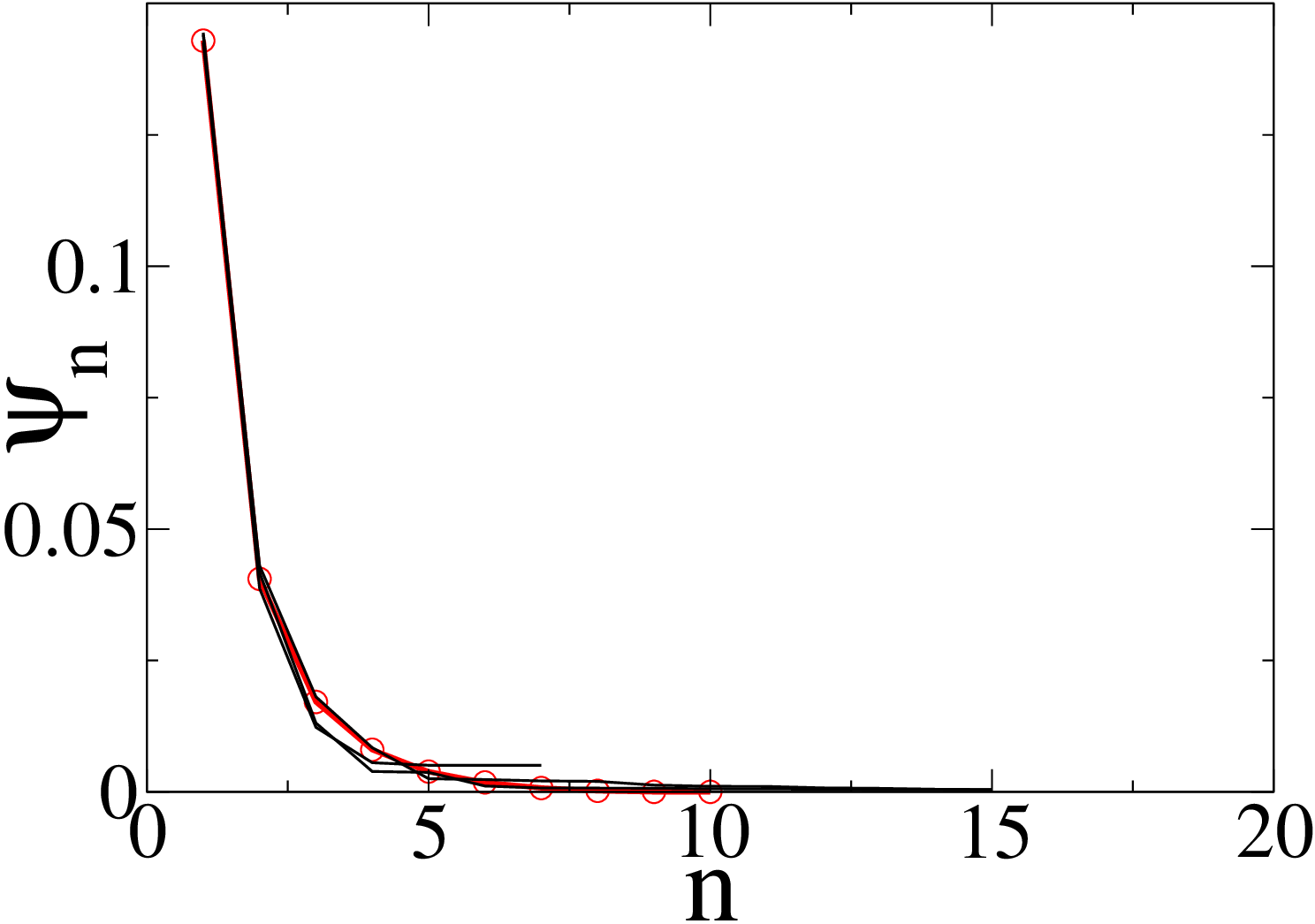}[\label{Fig333bbb}]
\end{Figure}

Fig.~\ref{Fig3} shows the size of the giant component in network A after each cascading step around the critical threshold for the first order phase transition (Fig.~\ref{Fig333aaa}) and the second order phase transition (Fig.~\ref{Fig333bbb}). One can see that simulation results of cascading failures agree well with analytical results, Eqs.~(\ref{Parshani1}) and~(\ref{Parshani2}). Different realizations give different results due to deviations from mean field, rendering small fluctuation around the mean field analytical results~\cite{Zhou2012}.\\

\begin{Figure}{\label{Fig4} Schematic representation of two types of NONs : (a) Star-like NON where one central network is interdependent with $(n-1)$ other networks. (b) Random regular NON where each network depends exactly on $m$ (here, $m=3$) other
networks. Circles represent interdependent networks and arrows represent interdependency relations. For example, $q_{12}$ represents a fraction $q_{12}$ of nodes in network 2 depend on nodes in network 1.}
\graphfile*[38]{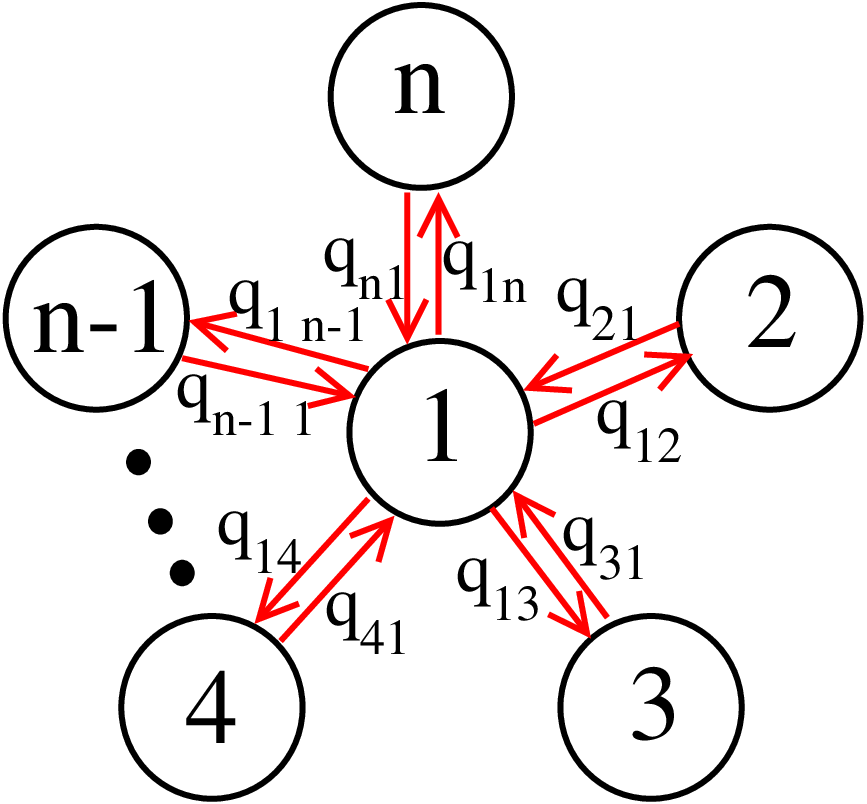}[\label{star}] 
\graphfile*[38]{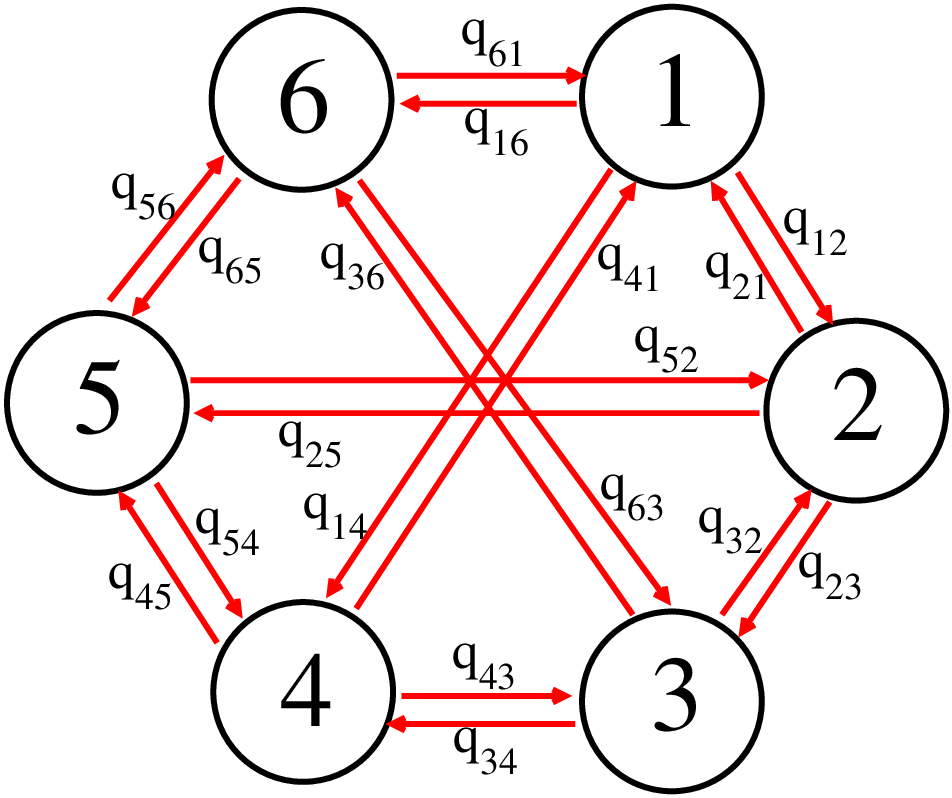}[\label{rrer}]
\end{Figure}

\section{Network of networks with clustering}
The framework discussed above can also be generalized to an interdependent system consisting of more than two networks. Here we consider two cases of NON~\cite{Gao2011,Gao2012, Gao2013} composed of $n$ interdependent networks, (i) A star-like NON and (ii) a random regular NON (see Fig.~\ref{Fig4}). We assume that for each pair of interdependent networks $i$ and $j$ ($i,j$=1,2,...,$n$), $q_{ji}$ denotes the fraction of nodes of network $i$ which depend on nodes of network $j$, i.e., they cannot function if the nodes upon which they depend fail. Similarly, $q_{ij}$ denotes the fraction of nodes nodes in network $j$ which depend on nodes of network $i$. After an initial attack of failure, only a fraction $p_i$ ($i=1,2,...,n$) of nodes in each network will remain. After  the process of cascading failures, a fraction $\psi_{\infty,i}$ of nodes in network $i$ will remain functional. The final giant component of each network can be express as $\psi_{\infty, i}=x_ig_i(x_i)$ and the unknowns $x_i$ can be found from a system of $n$ equations~\cite{Gao2011,Gao2012,Gao2013},
\begin{equation}
x_i=p_i\prod_{j=1}^K[q_{ji}y_{ji}g_j(x_j)-q_{ji}+1],\label{xi}
\end{equation}
where the product is taken over the K networks that are coupled with network $i$. By considering no-feedback condition~\cite{Gao2011,Gao2012,Gao2013}, we have 
\begin{equation}
y_{ji}=\frac{x_j}{q_{ij}y_{ij}g_i(x_i)-q_{ij}+1},\label{yi}
\end{equation}
which means the fraction of nodes left in network $j$ after damage from all networks coupled with network $j$ except network $i$.
Next we consider two analytically solvable examples for NON, the star-like network of ER networks and random regular (RR) network of ER networks, see Fig.~\ref{Fig4}.

\begin{figure}
   \includegraphics[width=0.48\textwidth, angle = 0]
   {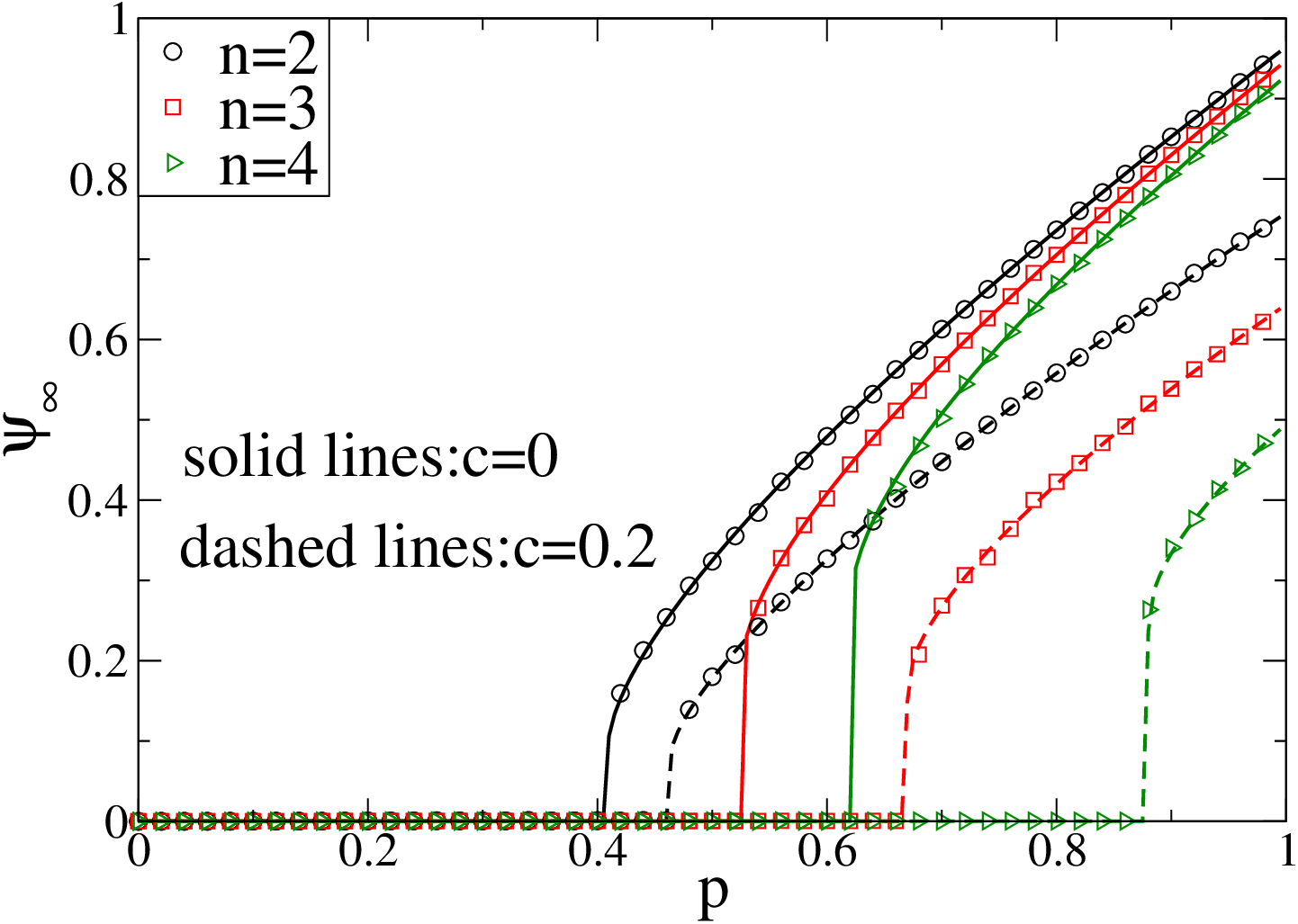}
\caption{Size of the giant component in the root network as a function of $p$ for $n=2, 3, 4$ and $c=0, 0.2$ for star-like NON. Average degree of each network in the NON is $\langle k \rangle$=4. Symbols and lines represent simulations ($N=10^5$) and theory, respectively.}
\label{Fig5}
\end{figure}

\subsection{Star-like NON with clustering}
For a star-like NON (Fig.~\ref{star}), we have a root network which is interdependent with other ($n-1$) networks. The initial attack is exerted for simplicity on the root network by removing a fraction of ($1-p$) of its nodes. The damage spreads to other networks, and comes back to the root network, back and forth. Here we consider the case for $n$ clustered ER networks with the same average degree $\langle k\rangle$ and same clustering coefficient $c$ (thus, the same average number of triangles $\langle t\rangle$). Assuming, for simplicity, that for all $i$, $q_{i1}=q_{1i}=q$, Eqs.~(\ref{xi}) and ~(\ref{yi}) are simplified to only two equations:
\begin{equation}
\begin{split}
x_1&=p[qg_2(x_2)-q+1]^{n-1},\\
x_2&=pqg_1(x_1)[qg_2(x_2)-q+1]^{n-2}-q+1. \label{Eq16}
\end{split}
\end{equation}
For clustered ER networks, $f(x)=1-g(x)$ satisfies 
\begin{equation}
f=exp[\langle t\rangle x^2(1-f)^2-\langle k\rangle x(1-f)]. \label{Eq17}
\end{equation}

By combining Eq.~(\ref{Eq16}) and Eq.~(\ref{Eq17}), we can find $x_1$, $x_2$ and $f_1$, $f_2$, from which the sizes of giant components in the root network ($\psi_{\infty}$) and other networks ($\phi_{\infty}$) can be obtained.

\begin{figure}
   \includegraphics[width=0.48\textwidth, angle = 0]
   {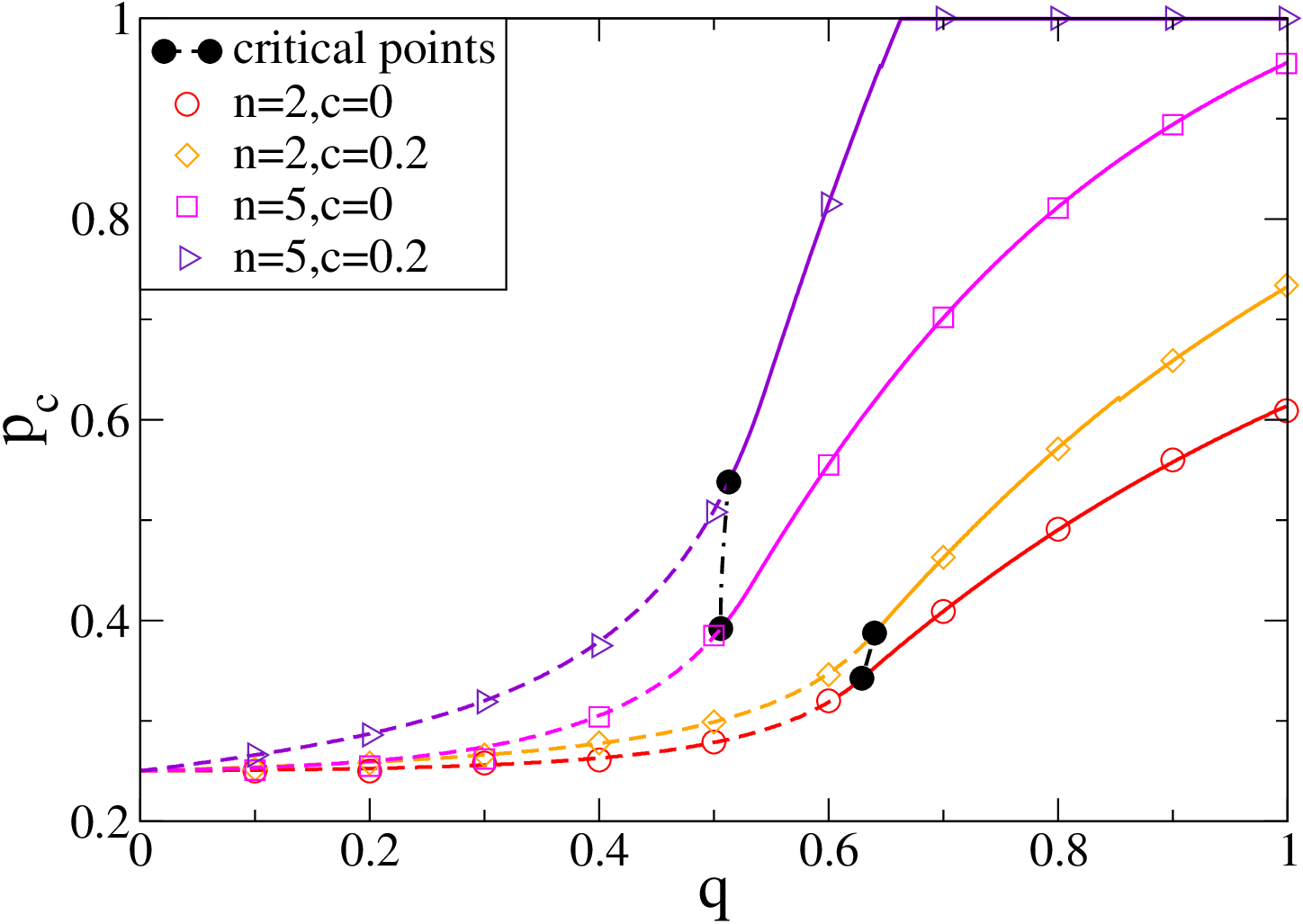}
\caption{Critical threshold $p_c$ as a function of interdependency strength $q$ for clustered star-like NON for $\langle k \rangle$=4, $n=2,5$ and $c=0, 0.2$. For each $n$ and $c$, there exsits a critical interdependency strength $q_c$ (solid symbols) that separates the first order (solid lines) and second order (dashed lines) phase transitions.}
\label{Fig6}
\end{figure}

Fig.~\ref{Fig5} shows the size of the giant component in the root network for $n=2,3,4$ comparing two cases $c=0$ (no clustering) and high clusteting, $c=0.2$. One can see that the simulation results agree well with theoretical results. Our results show that the NON becomes less robust with increasing $n$. For fixed $n$,  the NON composed of networks with larger clustering coefficient is less robust and the effect of clustering in reducing the robustness becomes larger as n increases. Similarly, the critical coupling $q_c$, where the behavior of phase transition changes from first order to second order decreases with n and slightly increases with clustering coefficient (see Fig.~\ref{Fig6}).

\subsection{Random regular (RR) NON of ER networks with clustering}
Now consider the case where each clustered ER network depends on exactly $m$ other clustered ER networks, i.e., random regular (RR) NON formed of clustered ER networks. Assume that the initial attack is exerted on each network with a fraction $(1-p)$ of randomly removed nodes and the interacting strengths are all equal to $q$. Furthermore, assume all ER networks have same average degree $\langle k\rangle$ and average number of triangles $\langle t\rangle$.  Now due to symmetry, all equations in Eqs.~(\ref{xi}) and ~(\ref{yi}) break down to a single equation and the size of giant component in each network is
\begin{equation}
\psi_{\infty}=p(1-e^{\langle t\rangle\psi^2_{\infty}-\langle k\rangle\psi_{\infty}})\Big[\frac{1-q+\sqrt{(1-q)^2+4q\psi_{\infty}}}{2}\Big]^m. \label{rrerphi}
\end{equation}

\begin{figure}
\includegraphics[width=0.48\textwidth, angle = 0]{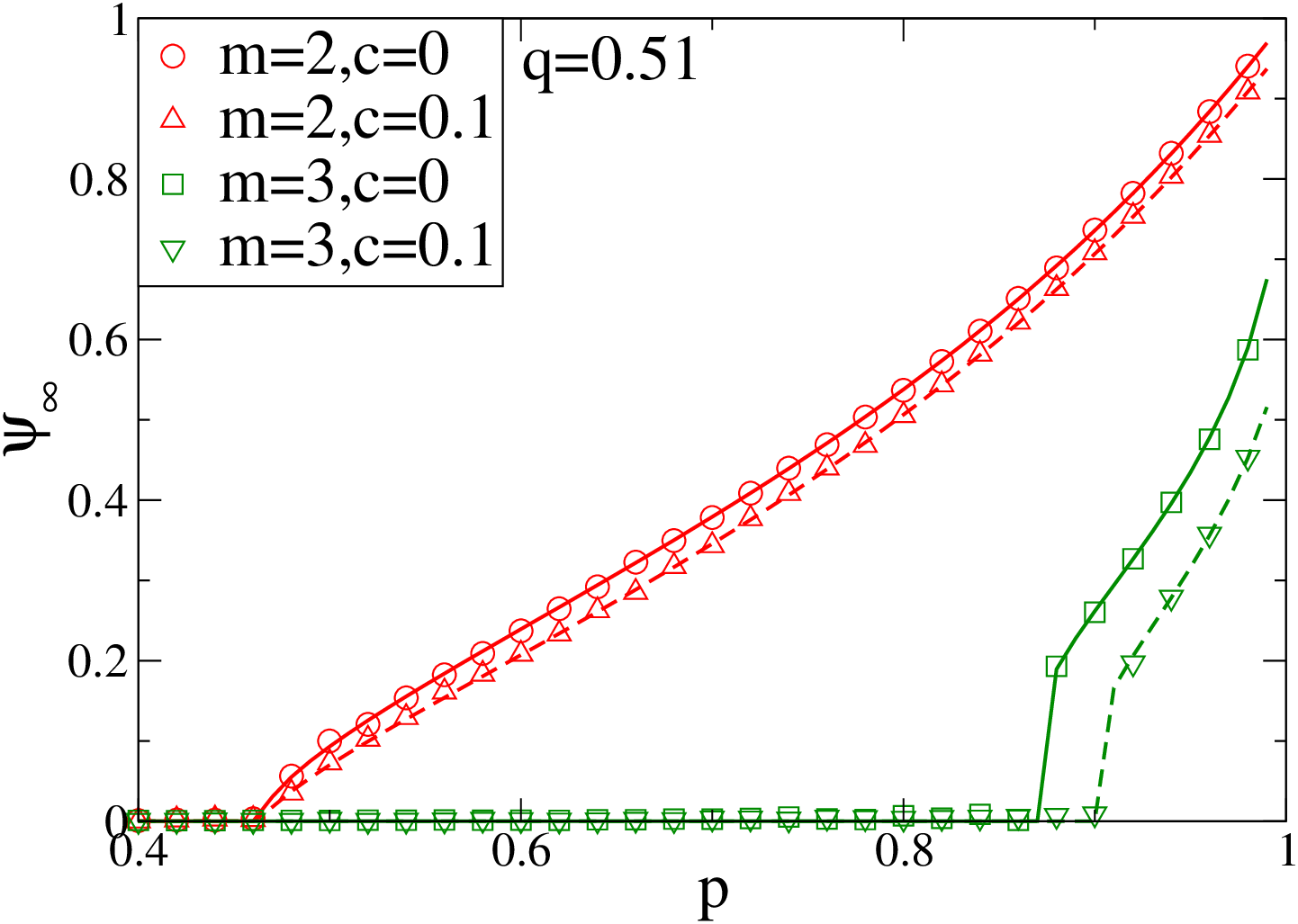}
\caption{Size of the giant component, $\psi_{\infty}$, as a function of $p$ for RR NON of clustered ER networks for fixed $q$ ($q$=0.51). The average degree is $\langle k \rangle$=9, $m$=2, 3 and $c$=0, 0.1. For $m=3$, the system shows a first order percolation transition as we change the value of $p$. While for $m=2$, the behavior of the phase transition is second order.}
\label{Fig7}
\end{figure}

\begin{figure}
   \includegraphics[width=0.48\textwidth, angle = 0]
   {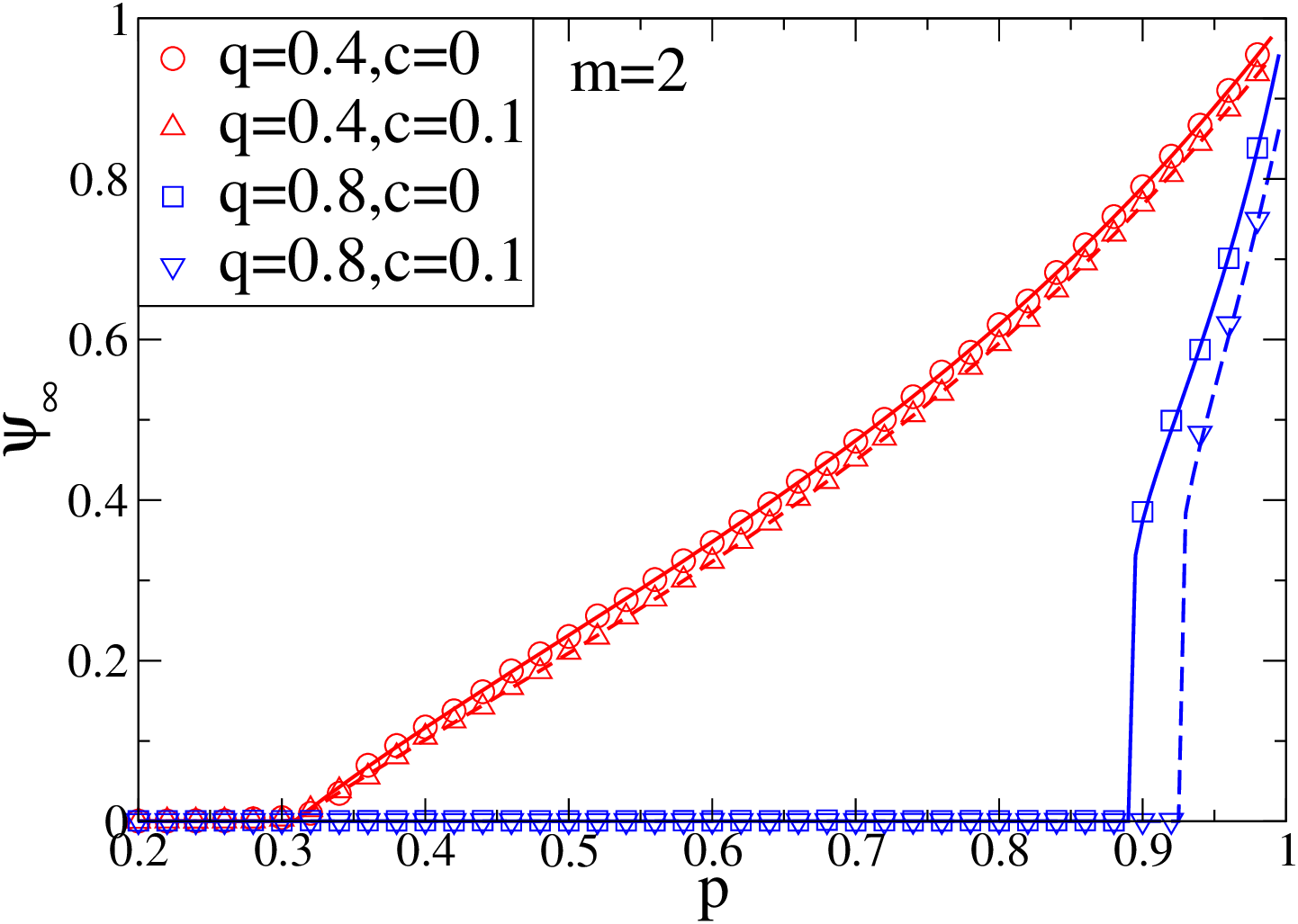}
\caption{Size of giant component, $\psi_{\infty}$, as a function of $p$ for RR NON composed of clustered ER networks for fixed $m$ ($m$=2). The average degree is $\langle k \rangle$=9, $q$=0.4, 0.8 and $c$=0, 0.1. The behavior of the phase transition is first order for $q=0.8$ and second order for $q=0.4$.}
\label{Fig8}
\end{figure}

Numerical solutions of Eq.~(\ref{rrerphi}) as well as simulation results are shown in Fig.~\ref{Fig7} and Fig.~\ref{Fig8}. One can see that the simulations agree very well with theory. For a given $\langle k\rangle$, the size of giant component  $\psi_{\infty}$ in each network shows a first order or second order phase transition as a function of $p$, depending on the values of $q$, $m$ and clustering coefficient $c$. As shown in Fig.~\ref{Fig7}, for some fixed values of $\langle k\rangle$ and $q$, the behavior of the phase transition can be first order or second order for different values of $m$. Similarly, as shown in Fig.~\ref{Fig8}, for fixed values of $\langle k\rangle$ and $m$, different values of $q$ can cause the phase transition to be first order or second order. In each scenario, for first order phase transtion, clustering within networks reduces the resistance of NON system to random nodes failure. While for second order phase transition, effect if clustering is similar but very small. This, again, is due to the smaller coupling $q$ in second order phase transition region.

By adding the condition that the first derivative of both sides of Eq.~(\ref{rrerphi}) with respect to $\psi_{\infty}$ are equal, we obtain the critical threshold of first order phase transition,  $p_{\uppercase\expandafter{\romannumeral1}}$. The critical threshold of second order phase transition $p_{\uppercase\expandafter{\romannumeral2}}$ is solved by adding condition $\psi_{\infty}(p_{\uppercase\expandafter{\romannumeral2}})\rightarrow 0$ to Eq.~(\ref{rrerphi}). By equating $p_{\uppercase\expandafter{\romannumeral1}}$ to $p_{\uppercase\expandafter{\romannumeral2}}$, the critical coupling $q_c$ where the first order phase transition changes to a second order phase transition can be solved analytically:
\begin{equation}
(\langle k\rangle^2+2\langle t\rangle)(1-q_c)^2=2\langle k\rangle q_cm.
\end{equation}
By substituting $c=\frac{2\langle t\rangle}{\langle k\rangle^2+2\langle t\rangle}$, we have
\begin{equation}
q_c=1+x-\sqrt{x(x+2)},\label{qcwithx}
\end{equation}
where $x\equiv \frac{m}{\langle k\rangle}(1-c)$. One can easily see that the increasing clustering coefficient $c$ will cause larger critical dependency $q_c$. Note that for $c=0$, Eq.~(\ref{qcwithx}) coincides with Eq. (30) in ~\cite{Gao2013}.

\begin{figure}
   \includegraphics[width=0.48\textwidth, angle = 0]
   {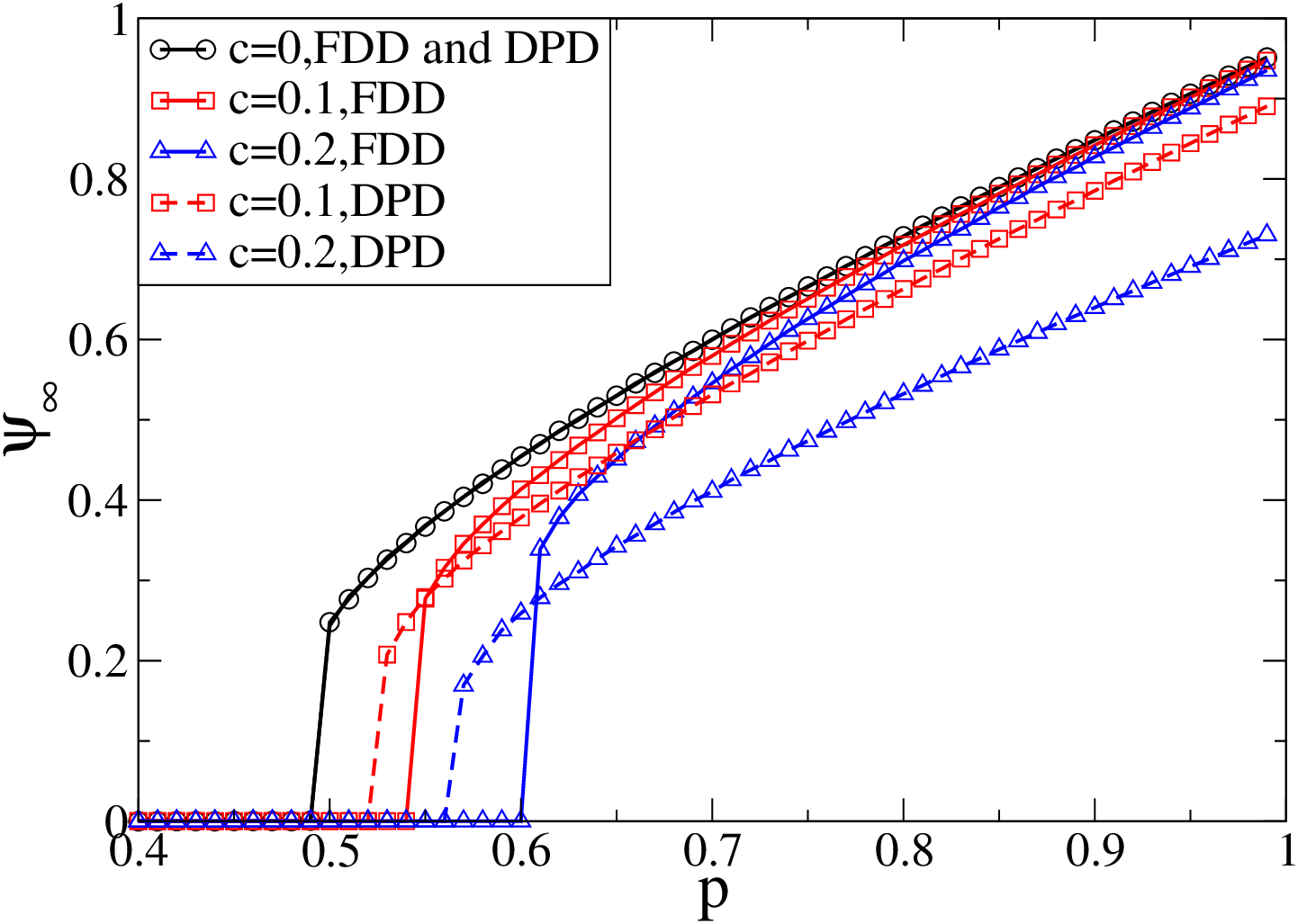}
\caption{Size of giant component in network A for two partially interdependent networks with clustering. Solid lines represent results for joint degree distribution which fix the total degree distribution being Poisson as we change the clustering coefficient (FDD). Dashed lines represent results for doubly Poisson distribution (DPD) with total average degree fixed. All resulted are from simulations with N=$10^6$, $\langle k\rangle$=4 and $q$=0.8. The behavior of the phase transition is first order.}
\label{Fig9}
\end{figure}

\section{Fixed Degree Distribution}
While doubly Poisson distribution model can seize the features of clustering and is possible to be solved analytically, the total degree distribution changes as the clustering coefficient changes for this model. Here we consider another kind of joint distribution $P_{st}$ proposed by Hackett et al.~\cite{Gleeson1,Gleeson2}, which preserves the total degree distribution $P(k)$ for different clustering coefficients. We set
\begin{equation}
P_{st}=P(k)\delta_{k,s+2t}[(1-f)\delta_{t,0}+f\delta_{t,\lfloor(s+2t)/2\rfloor}], \label{floor}
\end{equation}
where $f\in[0,1]$ and $\lfloor .\rfloor$ is the floor function~\cite{Gleeson1,Gleeson2}.\\
The above equation means that we construct $P_{st}$ from a given degree distribution $P(k)$ by picking a fraction $f$ of nodes being attached to a maximum possible number of triangles while the remaining $(1-f)$ nodes are attached to single edges only. From the definition of clustering coefficient, we have 
\begin{equation}
c=f\frac{\sum_kk(P(2k)+P(2k+1))}{\sum_k\binom{k}{2}P(k)},
\end{equation}
hence clustering coefficient can be adjusted by changing parameter $f$.\\
We investigate the effect of the joint degree distribution on robustness of partially interdependent networks by comparing the two joint degree distributions. One of which is fixed degree distribution (FDD), which is defined by Eq.~(\ref{floor}) with $P(k)$ obeying Poisson distribution ($P(k)$=$\langle k\rangle^k e^{-\langle k\rangle}/k!$). The other is the doubly Poisson distribution (DPD) studied in Sect. III, with $P_{st}$=$e^{-\langle s \rangle}\frac{\langle s \rangle^s}{s!}e^{-\langle t\rangle}\frac{\langle t\rangle^t}{t!}$.

\begin{figure}
   \includegraphics[width=0.48\textwidth, angle = 0]
   {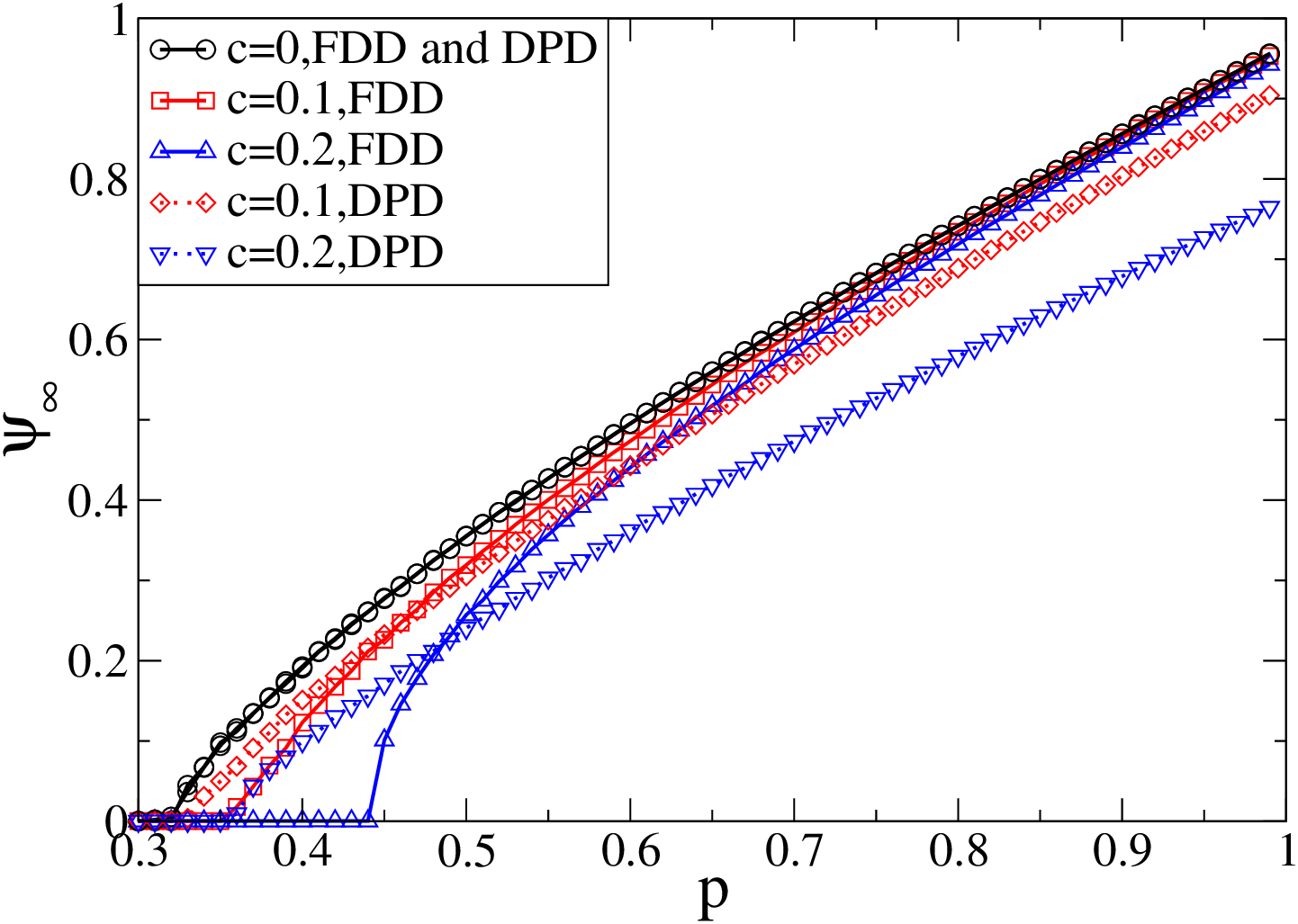}
\caption{Size of giant component in network A for two partially interdependent networks with clustering. Solid lines represent results for FDD and dashed lines represent results for DPD. All resulted are from simulations with N=$10^6$ and $\langle k\rangle$=4 and $q$=0.6. The behavior of the phase transition is second order.}
\label{Fig10}
\end{figure}

In Fig.~\ref{Fig9}, we plot the size of giant component in network A for two partially interdependent networks with clustering. The joint degree distribution in each network is chosen to be FDD and DPD, respectively. The interdependent strength $q$ is chosen to be that of the first order. One can see that the critical probability $p_c$ is larger for FDD compared with DPD with the same clustering coefficient. The differences in $p_c$ can be attributed to the broadening of $P(k)$ in the case of doubly Poisson distribution. Note that for site percolation on a sinlge clustered network, larger clustering coefficient leads to higher critical shreshold~\cite{Xuqing2012, Gleeson1}. Here for a system of two interdependent networks, the general trend is simliar, that for both degree distributions, $p_c$ increases as clustering coefficient being larger. The size of giant components for partially interdependent networks with second order phase transition for FDD and DPD are also shown in Fig.~\ref{Fig10}. The influence of clustering on the robustness of the partially interdependent networks is larger for FDD compared with DPD and the general trend is the same for two joint degree distributions.\\
\section{Conclusions}
We developed a framework for studying percolation of two partially interdependent ER networks with clustering. For each clustering coefficient, the system shows a first order to second order phase transition as we decrease coupling strength $q$. As we increase the clustering coefficient for each network, the system becomes less robust. This influence on robustness of network due to clustering coefficient becomes smaller as we decrease the coupling strength. Furthermore, the critical coupling strength $q_c$ at which first order phase transition changes to second order phase transition becomes larger as we increase the clustering coefficient.  This can be generalized to more than two networks. We considered two solvable cases for network of clustered networks (NONs) and found that higher clustering coefficient causes system to be less robust. Also, we investigated the difference and commonality for different joint degree distributions and found that though phase transition threshold is different from case to case, the general conclusion that increased clustering coefficient makes interdependent networks less robust holds for both cases.

\section*{Acknowledgments}
We wish to thank ONR (Grant N00014-09-1-0380, Grant N00014-12-1-0548), DTRA
(Grant HDTRA-1-10-1-0014, Grant HDTRA-1-09-1-0035), NSF (Grant CMMI 1125290),
the European projects LINC, MULTIPLEX (EU-FET project 317532), CONGAS (Grant FP7-ICT-2011-8-317672) and LINC, the DFG and the Israel Science Foundation for support.

\end{document}